\documentstyle[preprint,aps,eqsecnum]{revtex}
\begin{document}
\preprint{SNUTP 97-072, quant-ph/9706007}
\draft
\title{
Production of photons by the parametric resonance in the 
dynamical Casimir effect
}
\author{Jeong-Young Ji\footnote{Electronic address: jyji@phyb.snu.ac.kr}, 
Hyun-Hee Jung\footnote{Electronic address: hhjung@gmc.snu.ac.kr},
Jong-Woong Park\footnote{Electronic address: jwpark@gmc.snu.ac.kr}, 
Kwang-Sup Soh\footnote{Electronic address: kssoh@phya.snu.ac.kr}
}
\address{Department of Physics Education, Seoul National University, 
Seoul 151-742, Korea
}
\maketitle
\begin{abstract}
We calculate the number of photons produced by the parametric resonance 
in a cavity with vibrating walls. We consider the case that the frequency 
of vibrating wall is $n \omega_1 (n=1,2,3,...)$ which is a generalization 
of other works considering only $2 \omega_1$, where $\omega_1$ is the 
fundamental-mode frequency of the electromagnetic field in the cavity. For 
the calculation of time-evolution of quantum fields, we introduce a new 
method which is borrowed from the time-dependent perturbation theory of 
the usual quantum mechanics. This perturbation method makes it possible to 
calculate the photon number for any $n$ and to observe clearly the effect 
of the parametric resonance.
\end{abstract}


\narrowtext

\section{Introduction}

Recently, the photon creation in an empty cavity with moving boundaries, 
so called the dynamical Casimir 
effect~\cite{Name} 
has attracted much attention especially in a vibrating
cavity~\cite{Sassa94,Law94L,MeplanG96,LambrechtJR96}. 
It was also proposed that the high-$Q$ electromagnetic cavities may 
provide a possibility to detect the photons produced in the dynamical 
Casimir effect~\cite{Dodonov96,DodonovK96}. Therein, they considered the 
vibrating wall with the frequency $\Omega= 2 \omega_{1} $ and found the 
resonance excitation of the electromagnetic modes. 

For the solutions of Mathieu differential equation
\begin{equation}
\ddot{x} + \mu^{2} ( 1 + \epsilon \cos \Omega t) x = 0 ,
\label{Mathieu}
\end{equation} 
it is well known that parametric resonances occur when the frequency 
$\Omega $ with which the parameter oscillates is close to any value 
$2 \mu /n $ with $n$ integral~\cite{LandauL76T}. Since there are many 
mode frequencies for the differential equation describing the 
electromagnetic fields in a cavity, it is natural to consider other 
frequencies with $\Omega = n \omega_{1} . $ Now the method introduced in 
Ref.~\cite{DodonovK96} cannot be used in these general cases because the 
generating function can be made only for the special case $n=2 . $ 
Therefore we introduce a new perturbation method which leads to easily 
solvable equations. In principle, we may solve the equations to any order 
of expansion parameter $\epsilon,  $ but we calculate up to the first 
order of $\epsilon , $ which is sufficient to see the effect of the 
parametric resonance.

The organization of this paper is as follows. In Sec.~\ref{sec2} we 
review the scheme of the field quantization in the case of moving 
boundaries. In Sec.~\ref{sec3} we introduce the new perturbation method to 
find time evolution of quantum electromagnetic field. Here we write  the 
dominant part of the solution of wave equation which results from the 
parametric resonance. In Sec.~\ref{sec4} we calculate the number of 
photons created by the vibration of the boundary.
The last section is devoted to the summary and discussion. Here, we 
discuss the physical properties of the parametric resonance in the coupled 
differential equations, and we estimate the photon number created in 
realistic situation. Finally, the higher order calculations are considered 
briefly.

\section{Quantum electromagnetic fields in a cavity with moving walls}
\label{sec2} 

Let us consider an empty cavity formed by two perfect conducting walls, 
one being at rest at $x = 0 $ and the other moving according to a given 
law of motion $L(t) $ when $0 < t < T . $
The field operator in the Heisenberg representation $ A (x, t)  $ 
associated with a vector potential obeys the wave equation $(c =1) $
\begin{equation}
\frac{{{\partial^{2}}  A}}{\partial t^{2}} - \frac{{{ \partial^{2}}  
 A}}{\partial x^{2}}=0 
\label{WEQ}
\end{equation}
and can be written as
\begin{equation}
A (x,t) = \sum_{n} 
[b_{n} \psi_{n} (x,t) + 
b_{n}^{\dagger}\psi_{n}^{*}(x,t) ] .
\label{Aps}
\end{equation}
Here $b_{n}^{\dagger} $ and $b_{n}  $ are the creation and the 
annihilation operators and $\psi_{n} (x,t) $ is the corresponding mode 
function which satisfies the boundary condition 
$\psi_{n} (0, t) = 0 = \psi_{n} ( L(t), t) . $
 
For an arbitrary moment of time, following the approach of 
Refs.~\cite{Razavy,Calucci,Law94}, we expand the mode function as
\begin{equation}
\psi_{n} (x,t) = \sum_{k} Q_{nk} (t) \varphi_{k} (x,t)
\label{psn}
\end{equation}
with the instantaneous basis
\begin{equation}
\varphi_{k} (x, L(t)) = \sqrt{\frac{2}{L(t)}} \sin {\frac{{\pi k x 
 }}{L(t)} } .
\label{phn}
\end{equation}
Here $Q_{nk} (t) $ obeys an infinite set of coupled differential 
equations~\cite{Law95}:
\begin{eqnarray}
{\ddot{Q}}_{nk} + {\omega_{k} }^2 (t) Q_{nk}
&=&  2 \lambda \sum_{j} g_{kj} {\dot{Q}}_{nj}
+ \dot{\lambda} \sum_{j} g_{kj} Q_{nj} 
\nonumber \\
&& + \dot{\lambda}^{2} \sum_{j,l}  g_{jk} g_{jl} Q_{nl}
\label{EOM}
\end{eqnarray}
where $\lambda = \dot{L} / L $ and
\begin{equation}
g_{kj} = { \left\{\begin{array} {cc}(-1)^{k-j} {
\frac{{ 2 k j }}{ j^{2} - k^{2} } }~  & (j \neq k )  
\\ 0  & (j=k) \end{array}\right.} . 
\label{gkj}
\end{equation}
and the time-dependent mode frequency is
\begin{equation}
\omega_{k}  (t) = \frac{{ k \pi }}{L(t)} .
\label{omkt}
\end{equation}

For $t \leq 0 , $ the right hand side of Eq.~(\ref{EOM}) vanishes and the 
solution in this region is chosen to be
\begin{equation}
Q_{nk} (t) = 
\frac{{e^{-i \omega_{k} t}}}{\sqrt{2 \omega_{k}}}
\delta_{nk} .
\label{ICQnk}
\end{equation}
Then the quantum field can be written as
\begin{equation}
A (x, t \leq 0) = \sum_{n} 
[b_{n} \frac{{e^{-i \omega_{n} t}}}{\sqrt{2 \omega_{n}}} \varphi_{n} (x, 
 L_{0} ) 
+ {\rm H.c. }]
\end{equation}
$L_{0} $ is the initial distance between the walls and 
$\omega_{n} = \frac{{ \pi n }}{L_{0}} .
 $ 
From the form of the Hamiltonian 
$H = \sum_{n} \omega_{n} ( b_{n}^{\dagger} b_{n} + \frac{1}{2} )  ,
 $ 
we can interpret $b_{n}^{\dagger} b_{n} $ as the number operator 
associated with the photon with the frequency $\omega_{n} . $

After the oscillation of the wall $(t \geq T) , $ the solution of 
Eq.~(\ref{EOM}) with the initial condition (\ref{ICQnk}), can be written 
as
\begin{equation}
Q_{nk} (t \geq T) = \alpha_{nk} \frac{{e^{-i \omega_{k} t }}}{\sqrt{2 
 \omega_{k}}} +
\beta_{nk} \frac{{e^{i \omega_{k} t }}}{\sqrt{2 \omega_{k}}} . 
\label{QnkT}
\end{equation}
From (\ref{Aps}) and (\ref{psn}), we have
\begin{equation}
A (x, t \geq T) = \sum_{n} 
[a_{n} \frac{{e^{-i \omega_{n} t}}}{\sqrt{2 \omega_{n}}} \varphi_{n} (x, 
 L_{0} ) 
+ {\rm H.c. }] ,
\end{equation}
where the new creation and annihilation operators $a_{n}^{\dagger} $ and 
$a_{n} $ are written in terms of $b_{n}^{\dagger} $ and $b_{n} $, using 
the following Bogoliubov transformations
\begin{eqnarray}
a_{k} &=& \sum_{n} [b_{n} \alpha_{nk} + b_{n}^{\dagger} \beta_{nk}^* ]
\nonumber \\
a_{k}^{\dagger} &=& \sum_{n} [b_{n}^{\dagger} \alpha_{nk}^* 
+b_{n} \beta_{nk} ] . 
\label{a:b}
\end{eqnarray}
Further, it follows from 
$H = \sum_{n} \omega_{n} ( a_{n}^{\dagger} a_{n} + \frac{1}{2} ) 
 $ that $a_{n}^{\dagger} a_{n} $  is the new number operator at 
$t \geq T . $

If we start with a vacuum state $\left| 0_{b} \right> $ such that 
$b \left| 0_{b} \right> = 0 $, the expectation value of the new number 
operator is
\begin{equation}
N_{k} = \left<0_{b} \right| a_{k}^{\dagger} a_{k} \left| 0_{b} \right> 
= \sum_{n=1}^\infty  \left| \beta_{nk} \right|^{2}  ,
\label{NP} 
\end{equation}
which can be interpreted as the number of created photons. (One should 
note that the quantum state does not evolve in time in the Heisenberg 
picture.) 

\section{Time evolution of quantum electromagnetic fields
in a cavity with an oscillating wall}
\label{sec3}

In this section we find the time evolution of quantum field operator 
(\ref{Aps}) by solving Eq.~(\ref{EOM}) with the motion of the wall given 
by
\begin{equation}
L(t) = L_{0} [1 + \epsilon \sin(\Omega t)] .
\label{L}
\end{equation}
Here $\Omega = \gamma \omega_{1} = \gamma \pi / L_{0}  $ and $\epsilon $ 
is a small parameter characterized by the displacement of the wall. This 
is a generalization of the previous work in Ref.~\cite{DodonovK96} where 
the special case $\gamma=2 $ was treated. For $\epsilon \ll  1 , $ having 
in mind that $\lambda(t) \sim \epsilon $ and taking the first order of 
$\epsilon $ in the mode frequency (\ref{omkt})
\begin{equation}
\omega_{k} (t) = \frac{{ k \pi }}{L_{0}}
[1 + \epsilon \sin(\Omega t)]^{-1} ,
\label{om:e}
\end{equation}
we can replace Eq.~(\ref{EOM}) by a pair of coupled first-order 
differential equations
\begin{eqnarray}
{\dot{Q}}_{nk} &=& P_{nk} 
\nonumber \\
{\dot{P}}_{nk} &=& -{\omega_{k}}^2 ( 1 - 2 \epsilon \sin \Omega t) Q_{nk} 
+ 2 {\dot\frac{{L}}{L} } \sum_{j}  g_{kj} P_{nj} 
\nonumber \\ 
&& + \frac{{\ddot{L} }}{L} \sum_{j}  g_{kj} Q_{nj}  + O (\epsilon^{2} ).
\label{deQP}
\end{eqnarray}
Let us now introduce
\begin{equation}
X_{n, k \mp} = \sqrt{ \frac{\omega_{k}}{2}} 
\left( Q_{nk} \pm i \frac{P_{nk}}{\omega_{k}} \right)
\label{X:QP}
\end{equation}
or inversely,
\begin{eqnarray}
Q_{nk} &=& \sqrt{ \frac{1}{2 \omega_{k} }} [X_{n,k-} + X_{n,k+} ] 
\nonumber \\
P_{nk} &=& i \sqrt{ \frac{\omega_{k}}{2}} [ - X_{n,k-} + X_{n,k+} ].
\label{QP:X} 
\end{eqnarray}
Then (\ref{deQP}) reads
\begin{eqnarray}
{\dot{X}}_{n,k \mp} &=&  \mp i \omega_{k} X_{n,k-} 
\pm  i \omega_{k} \epsilon \sin \Omega t [ X_{n,k-} 
+ X_{n,k+} ] 
\nonumber \\
& \mp &  \epsilon \Omega \cos \Omega t \sum_{j} g_{kj} 
\sqrt{ \frac{\omega_{j}}{\omega_{k}} }
[- X_{n,j-} + X_{n,j+}] 
\nonumber \\
& \mp &  \frac{i}{2} \epsilon \Omega^{2} \sin \Omega t \sum_{j} g_{kj} 
{ \frac{1}{\sqrt{\omega_{j} \omega_{k}}}} [X_{n,j-} + X_{n,j+} ] .
\nonumber \\
&& \label{deX}
\end{eqnarray}

Introducing the infinite dimensional column vector
\begin{equation}
\vec{X}_{n}(t)  = { \left(\begin{array} {c}X_{n,1-} \\ X_{n,1+} \\ 
 X_{n,2-} \\  \vdots \end{array}\right)} ,
\end{equation}
the above equation can be written as a matrix form
\begin{equation}
\frac{d}{dt} \vec{X}_{n} (t) = V^{(0)} \vec{X}_{n}(t) 
+ \epsilon V^{(1)} \vec{X}_{n}(t) \label{deXn}
\end{equation}
and $V^{(0)} $ and $V^{(1)} $ are matrices.
The components of the matrices are
\begin{equation}
V_{k \sigma, j \sigma'}^{(0)} = i \omega_{k} \sigma \delta_{kj} 
 \delta_{\sigma \sigma'}
\end{equation}
and
\begin{equation}
V_{k \sigma, j \sigma'}^{(1)} = \sum_{s = \pm}
\omega_{1} v_{k \sigma, j \sigma'}^{s} e^{s i \gamma \omega_{1} t} ,
\label{Vv}
\end{equation}  
where
\begin{equation}
v_{k \sigma, j \sigma'}^{s} 
= \sigma \gamma g_{kj} \sqrt{\frac{j}{k}} 
\left( \frac{{ \sigma' }}{2} + s \frac{{\gamma}}{4j} \right)
- s \sigma \frac{k}{2} \delta_{kj}
\label{vpm}
\end{equation}
with $s, \sigma,\sigma' = +,-. $ Here we used $\Omega = \gamma \omega_{1} $ 
and $\omega_{k} = k \omega_{1} . $

To find the solution of Eq.~(\ref{deXn}), we introduce a perturbation 
expansion:
\begin{equation}
\vec{X}_{n} =\vec{X}_{n}^{(0)} + \epsilon \vec{X}_{n}^{(1)} 
+ \epsilon^{2} \vec{X}_{n}^{(2)} + \cdots .
\label{series}
\end{equation}
The iteration method used to solve this problem is similar to what we did 
in time-dependent perturbation theory. By inserting (\ref{series}) into 
Eq.~(\ref{deXn}), identifying powers of $\epsilon $ yields a series of 
equations: 
\begin{eqnarray}
\frac{d}{dt} \vec{X}_{n}^{(0)}
&=& V^{(0)} \vec{X}_{n}^{(0)}
\label{0th} \\
\frac{d}{dt} \vec{X}_{n}^{(1)}&=& V^{(1)} \vec{X}_{n}^{(0)} +V^{(0)} 
 \vec{X}_{n}^{(1)} .
\label{1st}
\end{eqnarray}
From the initial condition (\ref{ICQnk}), we have the solution to zeroth 
order equation (\ref{0th})
\begin{equation}
X_{n, k \sigma}^{(0)} = 
\delta_{nk} \delta_{\sigma-} e^{- i \omega_{k} t} .
\label{0thSol}
\end{equation}
Further, the Eq.~(\ref{1st}) is easily solved
\begin{equation}
X_{n, k \sigma}^{(1)} (t) = e^{\sigma i \omega_{k} t} 
\int_{0}^{t} dt' e^{- \sigma i \omega_{k} t' } \sum_{j, \sigma'} 
V_{k \sigma , j \sigma'}^{(1)}  X_{n, j \sigma'}^{(0)} .
\label{solX1}
\end{equation}
Using (\ref{0thSol}) and (\ref{Vv}), it can be written explicitly as
\begin{eqnarray}
X_{n, k \sigma}^{(1)} (t) 
&=& \omega_{1} e^{\sigma i k \omega_{1} t} 
\int_{0}^{t} dt' e^{- \sigma i k \omega_{1} t' } \sum_{j, \sigma'} 
( v_{k \sigma, j \sigma'}^{-} e^{-i \gamma \omega_{1} t'} 
\nonumber \\
&& + v_{k \sigma , j \sigma'}^{+} e^{i \gamma \omega_{1} t'} ) 
\delta_{nj} \delta_{- \sigma' } e^{- i j \omega_{1} t'} 
\nonumber \\
&=& \omega_{1} e^{\sigma i k \omega_{1} t} 
\int_{0}^{t} dt' 
( v_{k \sigma, n-}^{-} e^{-i(\sigma k + \gamma +n) \omega_{1} t'} 
\nonumber \\
&& + v_{k \sigma, n-}^{+} e^{+i(- \sigma k + \gamma -n) \omega_{1} t'} ) ,
\end{eqnarray}
that is, 
\begin{eqnarray}
X_{n, k +}^{(1)} (t) 
&=& 
- v_{k-,n-}^{-} E_{n + \gamma + k}^{-k} (t)
- v_{k-,n-}^{+} E_{n - \gamma + k}^{-k} (t) ,
\nonumber \\
X_{n, k -}^{(1)} (t) 
&=& 
v_{k-,n-}^{-} E_{n + \gamma - k}^{k} (t)
+ v_{k-,n-}^{+} E_{n - \gamma - k}^{k} (t) ,
\label{X1}
\end{eqnarray}
where 
\begin{equation}
E_{m}^{k} (t) = 
{ \left\{\begin{array} {cc}
\omega_{1} t e^{- i k \omega_{1} t}, & {\rm for} ~ m=0, \\
\frac{i}{ m } ( e^{- i (m+k) \omega_{1} t} - e^{-i k \omega_{1} t} ) , & 
 {\rm for} ~m \neq 0 .
\end{array}\right.}
\end{equation}
Therefore we have, using (\ref{X1}) and (\ref{QP:X}),
\begin{eqnarray}
Q_{nk} (t) &=&
 \frac{1}{\sqrt{2 \omega_{k}}} e^{-i \omega_{k} t} \delta_{nk} 
\nonumber \\
&+& \frac{\epsilon}{\sqrt{2 \omega_{k}}} [
- v_{k-,n-}^{-} E_{n + \gamma + k}^{-k} (t)
- v_{k-,n-}^{+} E_{n - \gamma + k}^{-k} (t) 
\nonumber \\
&+& v_{k-,n-}^{-} E_{n + \gamma - k}^{k} (t)
+ v_{k-,n-}^{+} E_{n - \gamma - k}^{k} (t) ]
\nonumber \\
&+& O( \epsilon^{2} ) .
\label{Qnk}
\end{eqnarray}
One should note that $Q_{nk}^{(1)} $ includes terms proportional to 
$\omega_{1} t $ which are the effects of parametric resonance. In the 
usual situation, since $\omega_{1} t \gg  1 $, only the resonance terms 
are dominant and the solution (\ref{Qnk}) becomes by retaining only them:
\begin{eqnarray}
Q_{nk} (t) & \approx &
 \frac{1}{\sqrt{2 \omega_{k}}} e^{-i \omega_{k} t} \delta_{nk} 
\nonumber \\
&+& \frac{{ \epsilon \omega_{1} t}}{\sqrt{2 \omega_{k}}} [
- v_{k-,n-}^{+} e^{i \omega_{k} t} \delta_{k, \gamma-n}
\nonumber \\
&+& v_{k-,n-}^{-} e^{-i \omega_{k} t} \delta_{k, n+\gamma}
+ v_{k-,n-}^{+} e^{-i \omega_{k} t} \delta_{k, n-\gamma} ] .
\nonumber \\
&&
\label{QnkReso}
\end{eqnarray}

\section{Number of photons created by the parametric resonance}
\label{sec4}

After some time interval $T$ the wall stops at $x=L_{0} $. Then the wave 
function becomes
\begin{equation}
\psi_{n} (x,t>T) = \sum_{k} \left[\alpha_{nk} \frac{{e^{-i \omega_{k} 
 t}}}{\sqrt{2 \omega_{k}}}
+\beta_{nk} \frac{{e^{i \omega_{k} t}}}{\sqrt{2 \omega_{k}}}\right] 
 \varphi_{k}(x) .
\label{psnT}
\end{equation}
The mode function (\ref{psn}) and its time-derivative should be 
continuous at $t=T : $
\begin{eqnarray}
\sum_{k} Q_{nk} (T) \varphi_{k}  = \sum_{k} \left( 
\alpha_{nk} \frac{e^{-i \omega_{k} T}}{\sqrt{2 \omega_{k}}}
+\beta_{nk} \frac{e^{i \omega_{k} T}}{\sqrt{2 \omega_{k}}}
\right) \varphi_{k}
\nonumber \\
\sum_{k} \left({\dot{Q}}_{nk}(T) \varphi_{k}
+ Q_{nk}(T) \dot{\varphi}_{k} \right) = 
~~~~~~~~~~~~~~~~~~~~~~~
\nonumber \\
~~~\sum_{k} \left( -i \omega_{k}
\alpha_{nk} \frac{e^{-i \omega_{k} T}}{\sqrt{2 \omega_{k}}}
+i \omega_{k} \beta_{nk} \frac{e^{i \omega_{k} T}}{\sqrt{2 \omega_{k}}}
\right) \varphi_{k} ,
\label {BC}
\end{eqnarray}
where we used (\ref{QnkT}) for the mode function at $t > T . $
By multiplying $\varphi_{l} $ to both sides of the above equations and 
integrating, we can get $\alpha_{nk~} {\rm and}~ \beta_{nk} $, which are 
\begin{eqnarray}
\alpha_{nk} &=& \left(i \omega_{k} Q_{nk} - {\dot{Q}}_{nk} 
+ {\dot\frac{{L}}{L}} \sum_{l} g_{kl} Q_{nl} \right) 
\frac{{e^{ i \omega_{k} T}}}{i \sqrt{2 \omega_{k} }}
\nonumber \\
\beta_{nk} &=&  \left(i \omega_{k} Q_{nk} + {\dot{Q}}_{nk} 
- {\dot\frac{{L}}{L}} \sum_{l} g_{kl} Q_{nl} \right) 
\frac{{e^{- i \omega_{k} T}}}{i \sqrt{2 \omega_{k} }} .
\label{EX}
\end{eqnarray}
Retaining only the dominant terms $(\omega_{1} T \gg  1) $
\begin{equation}
\beta_{nk} = - \epsilon \omega_{1} T v_{k-,n-}^+ \delta_{k, \gamma-n} 
e^{-i \omega_{k} T}.
\end{equation}
Using (\ref{gkj}) and (\ref{vpm}), finally we have
\begin{equation}
\left| \beta_{nk} \right|^{2}  =  \frac{1}{4} n k (\epsilon \omega_{1} 
 T)^{2} 
\delta_{k, \gamma-n }  .
\label{be2}         
\end{equation}
Therefore the total number of photons created in the $k $ th mode from 
the empty cavity is
\begin{equation}
N_{k} 
= \sum_{n=1}^\infty \left| \beta_{nk} \right|^{2}  
= { \left\{\begin{array}
{cc}{ \frac{1}{4} } (\gamma-k)k (\epsilon \omega_{1} T)^{2} 
 & k < \gamma \\
0, & {\rm otherwise} .
\end{array}\right.}
\label{Nk} 
\end{equation}
This result is a generalization of Ref.~\cite{DodonovK96} in the short 
time limit $(\epsilon \omega_{1} T \ll  1) $ and it agrees with that 
result for $\gamma =2  $ and $k =1 . $ It should also be noted that the 
maximal number of photons are created at the mode frequency
\begin{equation}
k = \frac{\gamma}{2} ~ {\rm or} ~ \omega_{k} = \frac{\Omega}{2} .
\label{most}
\end{equation}
for $\gamma = {\rm even} $ and at its nearest neighbor frequencies 
$k = (\gamma \pm 1)/2 $ for $\gamma = {\rm odd}. $

\section{Discussion}

We changed the second order coupled differential equation (\ref{EOM}) to 
the first order differential equation (\ref{deX}) by introducing the new 
variables (\ref{X:QP}). This makes it easy to deal with the differential 
equation and to find the perturbation series by virtue of diagonalization 
of $V^{(0)} . $ The $\epsilon^1$-order solution is found explicitly and it 
includes the terms proportional to time and this term is relatively large 
compared to other terms which include only oscillating parts. Considering 
only those dominant terms, we calculated the number of photons created 
after stopping of the wall vibration. The results show that the effect of 
parametric resonance is the largest at the half of the frequency of the 
vibrating wall $( \omega_{k} = \Omega/2 ) . $ This can be understood by 
considering the Mathieu equation (\ref{Mathieu}) where the parametric 
resonance takes place most strongly for $\Omega = 2 \mu . $ While, in the 
case $\gamma > 2 , $ we see the other resonance effects in addition to 
$\omega_{k} = \Omega/2 , $ which is due to the effect of couplings with 
other mode frequencies in the cavity. This can be interpreted as the 
parametric resonance for the coupled differential equation.

Following the discussions of Ref.~\cite{DodonovK96}, we estimate the rate 
of photon generation in the several modes. Our results (\ref{Nk}) which 
are valid only in the limit $\epsilon \omega_{1} T \ll  1 $ may be used to 
estimate the relative photon numbers depending on the mode frequency. For 
a long time, we assume that the photon numbers are proportional to 
time~\cite{LambrechtJR96,Dodonov96} or to exponential function of 
time~\cite{MeplanG96}. Then we have a photon {\it distribution} 
proportional to Eq.~(\ref{Nk}). For example, taking $\gamma = 4 $ we have  
$N_{2} = 4 N_{\rm D} $ and $N_{1} = N_{3} = 3 N_{\rm D} $ where $N_{k} $ 
is the number of created photons with $\omega_{k} . $ Here we used the 
same experimental parameters as considered in Ref.~\cite{DodonovK96}: 
$\epsilon_{\rm \max} \sim v_{s} \delta_{\rm \max} / \gamma \pi c \sim 1 
 \times 10^{-8} ,  $  $v_{s} \delta_{\rm \max} \sim 50 ~{\rm m/s} , $ 
$\omega_{1} / 2 \pi \sim 10 ~{\rm GHz} $ for $L_{0} \sim 2~{\rm cm} , $ 
and $N_{\rm D} \sim 300 ~{\rm photons} $ during $T = 1{\rm s}. $ For these 
experimental parameters, we expect that $N_{2} = 1200 ~{\rm photons} $ and 
$N_{1} = N_{3} = 900 ~{\rm photons}. $ 

The higher order calculations can be performed by taking the differential 
equation to higher order in $\epsilon $ and iterating the integral. Here 
we note that when we make the $\epsilon^p$-order differential equation 
like (\ref{deX}) from (\ref{EOM}) we consider the higher order coming from 
(\ref{om:e}). However if we limit the problem to the dominant terms which 
result from the parametric resonance, the problem is simple. Let us 
explain these by considering the $\epsilon^2$-order. The differential 
equation can be written, in the $\epsilon^2$-order, as
\begin{equation}
\frac{d}{dt} \vec{X}_{n}^{(2)} 
= V^{(2)} \vec{X}_{n}^{(0)}
+ V^{(1)} \vec{X}_{n}^{(1)} 
+ V^{(0)} \vec{X}_{n}^{(2)} .
\end{equation}
By integrating this differential equation it is clear that the $V^{(2)}$ 
term gives at most the term proportional to $\epsilon^{2} \omega_{1} T . $ 
This is very small compared to $ ( \epsilon \omega_{1} T )^{2} $ which 
comes from integrating the resonance term of $X^{(1)} . $ Therefore it is 
enough to write
\begin{equation}
\frac{d}{dt} \vec{X}_{n}^{(p)} 
= V^{(1)} \vec{X}_{n}^{(p-1)} 
+V^{(0)} \vec{X}_{n}^{(p)} ,
\end{equation}
to consider $p$-th order calculation. By iterations, we have 
\begin{eqnarray}
X_{n, k \sigma}^{(p)} (t) 
& \approx &
\frac{1}{p!} (\omega_{1} t)^{p} e^{\sigma i k \omega_{1} t} 
 \times
\nonumber \\
&& \sum_{\sigma_{p-1},...,\sigma_{1,} s_{p} ,...,s_{1}}
\delta_{0, - \sigma k + \Sigma_{p} \gamma - n} \times
\nonumber \\
&& v_{k, \sigma ; \sigma_{p-1} (\Sigma_{p-1} \gamma - n), 
\sigma_{p-1}}^{s_{p}}
\times
\nonumber \\
&& v_{\sigma_{p-1} (\Sigma_{p-1} \gamma - n), \sigma_{p-1};
\sigma_{p-2} (\Sigma_{p-2} \gamma - n), \sigma_{p-2}}^{s_{p-1}}
\times
\nonumber \\
&& \cdots  \times
\nonumber \\
&& v_{\sigma_{2} ( \Sigma_{2} \gamma - n), \sigma_{2} ; 
\sigma_{1} ( \Sigma_{1} \gamma - n), \sigma_{1}}^{s_{2}} 
\times
\nonumber \\
&& v_{\sigma_{1} ( \Sigma_{1} \gamma - n), \sigma_{1} ; n,-}^{s_{1}} ,
\end{eqnarray}
where $\Sigma_{k} = s_{1} + ... + s_{k} . $ 
For large $t$, we should be careful about the convergency of the series. 
We expect that the convergency property may be provided by the factor 
$1/p!$. However the explicit calculations are too difficult to find the 
simple general form of the solution and this will be our concern of future 
work~\cite{JiPrep}.

\section*{Acknowledgments}
This work was supported by the Center for Theoretical Physics (S.N.U.), 
Korea Research Center for Theoretical Physics and Chemistry, and the Basic 
Science Research Institute Program, Ministry of Education Project No. 
BSRI-96-2418. One of us (JYJ) is supported by Ministry of Education for 
the post-doctorial fellowship.

\end{document}